\documentclass[letterpaper,twocolumn,showpacs,prl,aps,nofootinbib,showkeys,superscriptaddress,10pt,nobalancelastpage]{revtex4-1}

\usepackage{graphicx}
\usepackage{amsmath}
\usepackage{mathtools}
\usepackage{amssymb}
\usepackage{hyperref}
\usepackage{color}
\usepackage{cleveref}
\usepackage{comment}

\crefname{figure}{Fig.}{Fig.}
\Crefname{figure}{Figure}{Figures}
\crefname{equation}{Eq.}{Eq.}
\Crefname{equation}{Equation}{Equations}
\crefname{section}{Sec.}{Sec.}
\Crefname{section}{Section}{Sections}
\creflabelformat{equation}{#2\textup{#1}#3}

\definecolor{darkgreen}{rgb}{0,.7,0}
\definecolor{linkblue}{rgb}{0.,0.,0.9333}

\makeatletter
\newcommand{\pushright}[1]{\ifmeasuring@#1\else\omit\hfill$\displaystyle#1$\fi\ignorespaces}
\newcommand{\pushleft}[1]{\ifmeasuring@#1\else\omit$\displaystyle#1$\hfill\fi\ignorespaces}
\makeatother

\hypersetup{
    pdffitwindow=true,      
    pdfnewwindow=true,      
    bookmarksnumbered=true,
    colorlinks=true,       
    linkcolor=red,          
    citecolor=darkgreen,        
    filecolor=magenta,      
    urlcolor=cyan,           
    menucolor=blue,
    breaklinks=true,
}

\begin{document}

\title{Even More Generalized Hamiltonian Dynamics}

\author{W.\ A.\ Horowitz}%
\email{wa.horowitz@uct.ac.za}
\affiliation{%
 Department of Physics, University of Cape Town, Private Bag X3, Rondebosch 7701, South Africa
}
\author{A.\ Rothkopf}%
\email{alexander.rothkopf@uis.no}
\affiliation{%
 Faculty of Science and Technology, University of Stavanger, 4021, Stavanger, Norway
}

\date{\today}

\begin{abstract}
    We establish the procedure to derive from an action-based variational principle the classical equations of motion in Hamiltonian phase space of a particle subject to general position and velocity dependent non-holonomic equality constraints.  Key to the procedure is our introduction 
    of Flannery brackets, which generalize Poisson brackets.  We conjecture on some implications, including the possibility of replacing Poisson brackets with Flannery brackets in Dirac's brackets to provide the quantization procedure for general non-holonomic equality constraint systems.  
\end{abstract}
\maketitle

\emph{Introduction} Since the publication of ``Generalized Hamiltonian Dynamics'' in 1950, Dirac's method of computing the equations of motion for constrained classical systems \cite{Dirac:1950pj} has been restricted to holonomic and integrable non-holonomic constraints.  By passing from the Lagrangian to the Hamiltonian formalism, Dirac's method provides an action-based independent alternative in phase space to the usual classical constrained Euler-Lagrange equations of motion in configuration space and forms the foundation of the quantization of constrained classical systems.  The most important examples of such constrained classical systems are gauge fields, and Dirac's method constitutes the basis of BRST quantization \cite{Henneaux:1992ig}.  We derive here a generalization of Dirac's method, thus providing for the first time in phase space an action-based derivation of the correct equations of motion for a system under general equality constraints. 

\emph{Dirac's Method for Holonomic and Integrable Non-holonomic Equality Constraints} Hamilton's Principle states that the classical path $q^i(t)$, $i=1,\ldots,n$, of an unconstrained particle extremizes the action
\begin{align}
    \label{eq:Ham}
    \delta_H S 
    & = \delta_H\int_{t_i}^{t_f} dt\,L(q^i,\dot q^i,t) \nonumber\\
    & = \int_{t_i}^{t_f} dt\, \left[ \frac{\partial L}{\partial q^i}\delta q^i + \frac{\partial L}{\partial \dot q^i}\delta\dot q^i \right]= 0,
\end{align}
where the variations of the path $\delta q^i(t)$ are unconstrained except at the boundaries $\delta q^i(t_i)=\delta q^i(t_f)=0$.  (Note the use of Einstein's summation convention above and throughout this Letter.)  That the variations are unconstrained and independent imply that 1) the transposition rule $\delta \dot q^i = (d/dt)\delta q^i$ holds and 2) one may use the fundamental theorem of the calculus of variations to derive the usual Euler-Lagrange equations of motion.

For particles subject to general equality constraints $g_k(q^i,\dot q^i,t)=0$, $k=1,\ldots,m$, Hamilton's Principle is modified to the Lagrange-d'Alembert Principle, which states that the classical path extremizes the action subject to variations in the path that are 1) zero at the boundaries, $\delta q^i(t_i)=\delta q^i(t_f)=0$, and 2) themselves consistent with the constraint equations, $g_k(q^i+\delta q^i,\dot q^i+\delta\dot q^i,t)=0$:
\begin{align}
    \label{eq:LdA}
    \delta_{LdA} S 
    & = \delta_{LdA}\int_{t_i}^{t_f} dt\,L(q^i,\dot q^i,t) \nonumber\\
    & = \int_{t_i}^{t_f} dt\, \left[ \frac{\partial L}{\partial q^i}\delta q^i + \frac{\partial L}{\partial \dot q^i}\delta\dot q^i \right]_{g_k(q^i+\delta q^i,\dot q^i+\delta\dot q^i,t)=0} \nonumber\\
    & = 0.
\end{align}
The Lagrange-d'Alembert Principle is much more difficult to utilize in practice than Hamilton's Principle: since the variations themselves are constrained, they are not all independent; one thus cannot straightforwardly apply the fundamental theorem of the calculus of variations to extract Euler-Lagrange-like equations of motion.  

For holonomic constraints, constraints that depend on the coordinates alone, $g_k(q^i)$, one may prove that adjoining the original Lagrangian with Lagrange multipliers multiplying the constraint equations, $\tilde L \equiv L - \lambda^k(t)g_k(q^i)$, then varying the adjoined action $\tilde S\equiv\int dt\tilde L$ with respect to all $q^i$ and $\lambda^k$, and treating all variations as independent and unconstrained, is equivalent to applying the Lagrange-d'Alembert Principle and leads to a set of Euler-Lagrange equations of motion.  

The equations of motion for general velocity dependent non-holonomic constraints are known \cite{Flannery:2011a}.  For velocity-dependent constraints, one can easily show that adjoining the Lagrangian as was done in the holonomic case and then extremizing the action with unconstrained variations yields incorrect equations of motion.  That simply adjoining the Lagrangian with velocity dependent constraints leads to the wrong equations of motion has caused signifcant confusion over the years \cite{Flannery:2005}, and it has been claimed that \emph{no} action extremization procedure can yield the correct equations of motion for non-linear in velocity non-holonomic constraints \cite{Flannery:2011a}.  

In \cite{Dirac:1950pj}, Dirac proposed a straightforward algorithm for finding the equations of motion for the classical path of a constrained particle in phase space.  The basic idea is to adjoin the Lagrangian with Lagrange multipliers multiplying the constraints, pass from the Lagrangian to the Hamiltonian, and then use the arbitrariness of the time evolution of the Lagrange multipliers to ensure that the system never leaves the constraint surface in phase space.  To wit, to employ Dirac's method, one takes $\tilde L \equiv L - \lambda^k(t)g_k(q^i,\dot q^i,t)$.  Then the adjoined Hamiltonian is
\begin{align}
    \tilde H \equiv \dot q^i p_i + \dot \lambda^k p_{\lambda_k} - \tilde L, \qquad p_\alpha \equiv \frac{\partial\tilde L}{\partial \dot q^\alpha}
\end{align}

Note that, formally, $p_{\lambda_k}\equiv0$ because $\tilde L$ doesn't depend on $\dot\lambda^k$.  Nevertheless, we may formally retain the $p_{\lambda_k}$ dependence.  If we assume, as implicitly done by Dirac, that the transposition rule $\delta \dot q^i = (d/dt)\delta q^i$ holds for the variations, then
\begin{align}
    \delta \tilde S
    & = \int dt \biggl[ p_i\delta \dot q^i + \dot q^i\delta p_i + p_{\lambda_k} \delta \dot \lambda^k + \dot\lambda^k\delta p_{\lambda_k} \nonumber\\
    & \qquad - \frac{\partial \tilde H}{\partial q^i}\delta q^i - \frac{\partial \tilde H}{\partial p_i}\delta p_i - \frac{\partial \tilde H}{\partial \lambda^k}\delta\lambda^k + \frac{\partial\tilde H}{\partial p_{\lambda_k}}\delta p_{\lambda_k}\biggr] \nonumber\\[5pt]
    & = \int dt \Biggl[ -\left(\dot p_i + \frac{\partial\tilde H}{\partial q^i}\right)\delta q^i + \left( \dot q^i - \frac{\partial\tilde H}{\partial p_i} \right)\delta p_i \nonumber\\
    & \qquad - \left(\dot p_{\lambda_k} + \frac{\partial\tilde H}{\partial \lambda^k}\right)\delta \lambda^k + \left( \dot \lambda^k - \frac{\partial\tilde H}{\partial p_{\lambda_k}} \right)\delta p_{\lambda_k}\Biggr] \nonumber\\[5pt]
    & = 0.
\end{align}
If we \emph{assume} that the variations are all independent, we arrive at the extended Hamilton's equations
\begin{align}
    \label{eq:Dirac}
    \begin{split}
        \dot q^i & = \frac{\partial\tilde H}{\partial p_i} \\
        \dot \lambda^k & = \frac{\partial\tilde H}{\partial p_{\lambda_k}} 
    \end{split}
    &
    \begin{split}
        \dot p_i & = -\frac{\partial\tilde H}{\partial q^i} \\
        \dot p_{\lambda_k} & = -\frac{\partial\tilde H}{\partial \lambda^k}.
    \end{split}
\end{align}
Extending the coordinates $q^i$ and their conjugate momenta $p_i$ to include the Lagrange multipliers and their conjugate momenta then yields extended Poisson brackets
\begin{align}
    \{X,Y\}_{PB}\equiv\frac{\partial X}{\partial q^i}\frac{\partial Y}{\partial p_i} - \frac{\partial X}{\partial p_i}\frac{\partial Y}{\partial q^i} + \frac{\partial X}{\partial \lambda^k}\frac{\partial Y}{\partial p_{\lambda_k}} - \frac{\partial X}{\partial p_{\lambda_k}}\frac{\partial Y}{\partial \lambda^k}
\end{align}
that gives the time evolution according to \cref{eq:Dirac} for any function $X(q^i,p_i,\lambda^k,p_{\lambda_k})$,
\begin{align}
    \frac{dX}{dt} = \{X,\tilde H\}_{PB}.
\end{align}
In Dirac's method, one substitutes undetermined functions $u_{\lambda_k}(q^i,p_i,\lambda^k)$ for the $\dot\lambda^k$ in the adjoined Hamiltonian; we are free to make this substitution because the $p_{\lambda_k}\equiv0$.  We then compute the time evolution of the $p_{\lambda_k}$ according to \cref{eq:Dirac} and adjust the $u_{\lambda_k}$ to ensure that time evolution does not evolve the $p_{\lambda_k}$ away from 0.  One may readily confirm the validity of Dirac's method against simple examples of holonomic and integrable constraints, e.g.\ for the rigid rod pendulum. 

\begin{figure}[!tbp]
    \centering
    \includegraphics[width=\columnwidth]{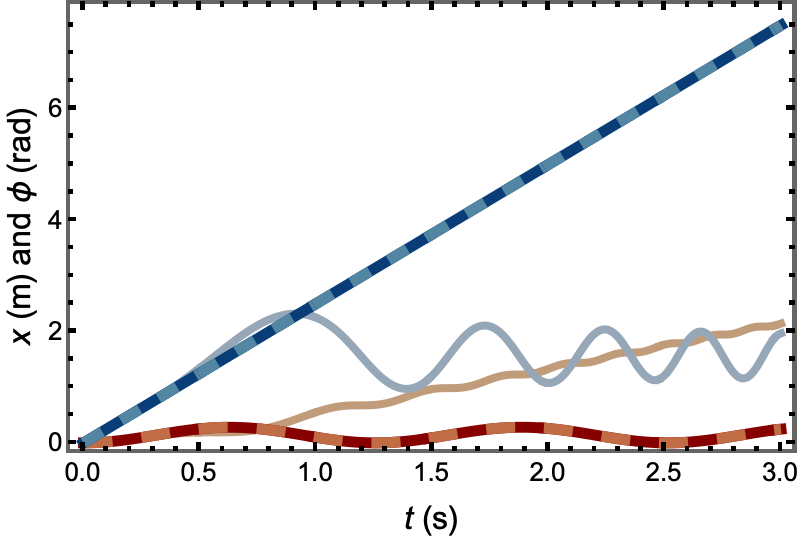}
    \caption{
    Plots of $x(t)$ (reds) and $\phi(t)$ (blues) for a 1 kg uniform sphere of radius 1 m rolling without slipping in a uniform gravitational field with $g=9.8$ m/s$^2$ down a slope of angle $\alpha=\pi/6$ with intial velocities $\dot x=\dot y = \dot\theta=0$ and $\dot\phi=2.5$ 1/s solved from the known non-holonomic equations of motion \protect\cite{Flannery:2011a} (darkest), Dirac's original method \protect\cite{Dirac:1950pj} (lightest), and our extension of Dirac's method to general velocity dependent non-holonomic constraints (dashed).  $x$ is the distance travelled by the sphere down the slope and $\phi$ is the angle of rotation about the sphere's vertical axis.
    }
    \label{fig:motion}
\end{figure}

\emph{General Equality Constraints} We show in \cref{fig:motion} the dramatic breakdown of the usual Dirac method (solid lightest colors) when applied to the canonical non-integrable constraint problem of rolling without slipping.  Perhaps most shocking, one can see that the original Dirac method completely misses the physics that the angular velocity of the rotation of the sphere about its vertical axis is a constant.  

The key to understanding the breakdown of Dirac's method was recently illuminated by Flannery \cite{Flannery:2011a}.  What Flannery realized is that the transposition rule is in general
\begin{align}
    \frac{\partial g_k}{\partial \dot q^i}\left[ \delta\dot q^i - \frac{d}{dt}\delta q^i \right] = g_{kj}\delta q^j, \, g_{kj}\equiv\frac{d}{dt}\frac{\partial g_k}{\partial \dot q^j} - \frac{\partial g_k}{\partial q^j},
\end{align}
which reduces to the usual transposition rule $\delta\dot q^i=(d/dt)\delta q^i$ for holonomic or integrable non-holonomic constraints, since $\partial g_k/\partial \dot q^i\equiv0$ and $g_{kj}\equiv0$ in these cases, respectively.  
If we define $f^i_j$
\begin{align}
    \delta\dot q^j - \frac{d}{dt}\delta q^j \equiv f^j_{\phantom ji} \delta q^i,
\end{align}
then the variation of the $\tilde S$ action above is modified to
\begin{align}
    \delta \tilde S
    & = \int dt \biggl[ p_i\delta \dot q^i + \dot q^i\delta p_i + p_{\lambda_k} \delta \dot \lambda^k + \dot\lambda^k\delta p_{\lambda_k} \nonumber\\
    & \qquad - \frac{\partial \tilde H}{\partial q^i}\delta q^i - \frac{\partial \tilde H}{\partial p_i}\delta p_i - \frac{\partial \tilde H}{\partial \lambda^k}\delta\lambda^k + \frac{\partial\tilde H}{\partial p_{\lambda_k}}\delta p_{\lambda_k}\biggr] \nonumber\\[5pt]
    & = \int dt \Biggl[ -\left(\dot p_i + \frac{\partial\tilde H}{\partial q^i} - p_jf^j_{\phantom ji}\right)\delta q^i + \left( \dot q^i - \frac{\partial\tilde H}{\partial p_i} \right)\delta p_i \nonumber\\
    & \qquad - \left(\dot p_{\lambda_k} + \frac{\partial\tilde H}{\partial \lambda^k}\right)\delta \lambda^k + \left( \dot \lambda^k - \frac{\partial\tilde H}{\partial p_{\lambda_k}} \right)\delta p_{\lambda_k}\Biggr] \nonumber\\[5pt]
    & = 0.
\end{align}
Since the variations of the Lagrange multipliers are unconstrained, the transposition rule continues to hold for them, $\delta\dot\lambda^k = (d/dt)\delta\lambda^k$.  As done by Dirac, we assume that all the variations are now independent and are thus led to the modified extended Hamilton's equations
\begin{align}
    \label{eq:Flannery}
    \begin{split}
    \dot q^i & = \frac{\partial\tilde H}{\partial p_i} \\
    \dot \lambda^k & = \frac{\partial\tilde H}{\partial p_{\lambda_k}}
    \end{split}
    &
    \begin{split}
        \dot p_i & = -\frac{\partial\tilde H}{\partial q^i} + p_jf^j_{\phantom ji} \\
        \dot p_{\lambda_k} & = -\frac{\partial\tilde H}{\partial \lambda^k}.
    \end{split}
\end{align}
If we now define \emph{Flannery brackets} for generalized coordinates $q^i$ and momenta $p_i$ (which may be extended to include Lagrange multipliers and their conjugate momenta as was done above) by
\begin{align}
    \{X,Y\}_{FB}\equiv\frac{\partial X}{\partial q^i}\frac{\partial Y}{\partial p_i} - \frac{\partial X}{\partial p_i}\left(\frac{\partial Y}{\partial q^i}-p_jf^j_{\phantom ji}\right), 
\end{align}
then time evolution according to \cref{eq:Flannery} is given by
\begin{align}
    \frac{dX}{dt} = \{X,\tilde H\}_{FB}.
\end{align}
If one now follows Dirac's procedure, but with the replacement of Poisson brackets with Flannery brackets, then one may successfully determine the equations of motion for general equality constraints.  

Note that for holonomic or for integrable non-holonomic constraints $f^j_{\phantom ji}\equiv0$ \cite{Flannery:2011a}.  In these cases the Flannery brackets are the same as the Poisson brackets, and it's clear that the method presented here is a generalization of Dirac's original method that gives exactly the same results as Dirac's original method for holonomic or integrable non-holonomic constraints.

We show in \cref{fig:motion} the success of our generalization of Dirac's method to the particular non-integrable constraint problem of rolling without slipping.  

\emph{Conclusions and Outlook} Dirac's method of deriving the equations of motion for constrained classical systems is restricted to holonomic and integrable non-holonomic constraints.  We derived here a generalization of Dirac's method that applies to holonomic and fully general in velocity non-holonomic equality constraints.  To do so we introduced an extension of the usual Poisson brackets, which we denote Flannery brackets, that take into account the modification of the transposition rule relating $\delta\dot q^i$ and $(d/dt)\delta q^i$ induced by general non-holonomic equality constraints.  

There are many future important avenues of investigation that we leave to future pursuits. These include investigating: non-holonomic equality constraints that depend on accelerations \cite{Flannery:2011a}; whether the inverse Legendre transform of the modified Hamiltonian $\tilde H$ yields a Lagrangian whose variation gives the correct non-holonomic equations of motion; if Flannery brackets obey a Liouville-type theorem; implications for conservation laws and Noether's theorem in general and the conservation of the energy of the system in particular, since Flannery brackets are not automatically antisymmetric.

Currently there is no general theory for quantizing general non-holonomically constrained classical systems \cite{Fernandez:2018}.  Since Dirac brackets, which are Poisson brackets properly restricted to the constraint surface in phase space, are the foundation for quantizing systems of integrable constraints, it may be possible to replace the Poisson brackets with Flannery brackets in the Dirac brackets to find such a theory for quantizing systems with general non-holonomic constraints.  It may also be possible to modify Moyal brackets with Flannery brackets.  We note, however, that the Flannery brackets introduced here do not satisfy the same properties as Poisson brackets, which may complicate such a quantization procedure.  Alternatively, Flannery brackets may provide insight into possible failures of quantization in generally non-holonomic equality constraint systems.

\emph{Acknowledgements} WAH thanks the National Research Foundation and SA-CERN Collaboration for financial support.  AR and WAH acknowledge support from the ERASMUS+ project 2023-1-NO01-KA171-HED-000132068.  The authors also thank M.\ Henneaux for valuable comments.

\providecommand{\newblock}{}

\end{document}